\documentclass[a4paper, conference]{IEEEtran}
\usepackage[T1]{fontenc}
\usepackage[utf8]{inputenc}
\usepackage{floatrow} 
\usepackage{subfig}
\usepackage{amsmath}
\usepackage{amssymb}
\usepackage[super]{nth}
\usepackage{paralist}
\usepackage{pgfplots}
\usepackage{tikz}
\usepackage{algorithm}
\usepackage[noend]{algpseudocode}
\usepackage{textcomp}

\algnewcommand\algorithmicforeach{\textbf{for each}}
\algdef{S}[FOR]{ForEach}[1]{\algorithmicforeach\ #1\ \algorithmicdo}

\usetikzlibrary{calc}
\usetikzlibrary{intersections}
\usetikzlibrary{positioning}
\usetikzlibrary{shapes}
\usetikzlibrary{snakes}

\tikzstyle{circ} = [shape = circle, draw, inner sep = 1 pt]
\tikzstyle{circ1} = [circ]
\tikzstyle{circ2} = [circ, fill = black!30]
\tikzstyle{icwave} = [->, > = stealth, red, decorate,
  decoration = {snake, segment length = 2 mm, post length = 1 mm}]
\tikzstyle{muwave} = [<->, > = stealth, gray, decorate,
  decoration = {snake, amplitude = 0.25 mm, segment length = 1 mm,
    post length = 1 mm, pre length = 1 mm}]
\tikzstyle{iclink} = [->, > = stealth, red]
\tikzstyle{mycloud} = [cloud, cloud puffs = 15, cloud ignores
  aspect, minimum width = 3 cm, minimum height = 1.25 cm, align = center,
  draw]
\tikzstyle{switch} = [draw, inner sep = 0, minimum width = 0.5 cm,
  minimum height = 0.5 cm]
\tikzstyle{router} = [draw, circle, inner sep = 0, minimum size = 0.5 cm]
\tikzstyle{fmn} = [draw, ellipse, align = center, minimum width = 1.5 cm]
\tikzstyle{arn} = [circle, draw, inner sep = 1.5 pt]
\tikzstyle{prn} = [arn, fill = black]
\tikzstyle{sprn} = [draw, inner sep = 1.5 pt, fill = black]

\newcommand{\BS}[1]{%
  \begin{tikzpicture}[#1]
    \fill (0, 0) circle (5 mm);
    \draw ([shift = (-45:15 mm)] 0, 0) arc (-45:45:15 mm);
    \draw ([shift = (135:15 mm)] 0, 0) arc (135:225:15 mm);
    \draw ([shift = (-45:10 mm)] 0, 0) arc (-45:45:10 mm);
    \draw ([shift = (135:10 mm)] 0, 0) arc (135:225:10 mm);
    \coordinate (ll) at ($ (0, 0) + (-15 mm, -40 mm) $);
    \coordinate (rl) at ($ (0, 0) + (15 mm, -40 mm) $);
    \draw (0, 0) -- (ll) -- ($ (0, 0)!0.5!(rl) $) -- ($ (0, 0)!0.25!(ll) $);
    \draw (0, 0) -- (rl) -- ($ (0, 0)!0.5!(ll) $) -- ($ (0, 0)!0.25!(rl) $);
  \end{tikzpicture}%
}

\newcommand{\FU}[1]{%
  \begin{tikzpicture}[#1]
    \draw (-20 mm, -15 mm) rectangle (20 mm, 15 mm);
    \draw (-15 mm, -5 mm) rectangle (0 mm, 10 mm);
    \draw (-25 mm, 13 mm) -- (0, 30 mm) -- (25 mm, 13 mm);
    \draw (5 mm, -15 mm) -- (5 mm, 10 mm) -- (15 mm, 10 mm) -- (15 mm, -15 mm);
  \end{tikzpicture}%
}

\newcommand{\MU}[1]{%
  \begin{tikzpicture}[#1]
    \draw (3 mm, 0) arc (0:90:3 mm);
    \draw (5 mm, 0) arc (0:90:5 mm);
    \draw (-10 mm, -15 mm) rectangle (0 mm, 0 mm);
    \draw (-8 mm, -13 mm) rectangle (-2 mm, -2 mm);
  \end{tikzpicture}%
}

\newcommand{\VDOTS}[1]{%
  \begin{tikzpicture}[#1]
    \fill [black] ($ (0, 0) + (0, 4 pt) $) circle (1 pt);
    \fill [black] (0, 0) circle (1 pt);
    \fill [black] ($ (0, 0) + (0, -4 pt) $) circle (1 pt);
  \end{tikzpicture}%
}

\newcommand{\sprn}{%
  \begin{tikzpicture}
    \node [sprn] (0, 0) {};
  \end{tikzpicture}%
}

\newcommand{\RP}{%
  \begin{tikzpicture}

    \draw [black, dash pattern = on 0 pt off 5 pt on 10 pt off 5 pt,
    line width = 3 pt, opacity = 0.5, rounded corners = 4 pt, line cap
    = round]
    (0, 0) -- (0.75, 0);

    \draw [green, dash pattern = on 0 pt off 5 pt on 10 pt off 5 pt,
    line width = 2.5 pt, opacity = 0.5, rounded corners = 4 pt, line
    cap = round]
    (0, 0) -- (0.75, 0);

  \end{tikzpicture}%
}

\newcommand{\WP}{%
  \begin{tikzpicture}

    \draw [black, dash pattern = on 6 pt off 5.5 pt, line width = 3 pt,
    opacity = 0.5, rounded corners = 4 pt, line cap = round]
    (0, 0) -- (0.6, 0);

    \draw [red, dash pattern = on 6 pt off 5.5 pt, line width = 2.5
    pt, opacity = 0.5, rounded corners = 4 pt, line cap = round]
    (0, 0) -- (0.6, 0);

  \end{tikzpicture}%
}

\title{Performance of Interoperator\\Fixed-Mobile Network Sharing}

\author{

  \IEEEauthorblockN{Ireneusz Szcześniak\IEEEauthorrefmark{1}, Andrzej
    R.~Pach\IEEEauthorrefmark{2} and Bożena
    Woźna-Szcześniak\IEEEauthorrefmark{3}}

  \IEEEauthorblockA{\IEEEauthorrefmark{1}Częstochowa University of
    Technology, Institute of Computer and Information
    Sciences\\ul.~J.~H.~Dąbrowskiego 73, 42-201 Częstochowa, Poland}

  \IEEEauthorblockA{\IEEEauthorrefmark{2}AGH University of Science and
    Technology, Department of Telecommunications\\al.~Mickiewicza 30,
    30-059 Kraków, Poland}

  \IEEEauthorblockA{\IEEEauthorrefmark{3}Jan Długosz University in
    Częstochowa, Institute of Mathematics and Computer
    Science\\al.~Armii Krajowej 13/15, 42-200 Częstochowa, Poland}

}

\begin{document}

\maketitle


\begin{abstract}
  We evaluate the downstream performance of our novel interoperator
  fixed-mobile network sharing, in which operators exchange data in
  their access networks.  We propose a performance evaluation
  algorithm, and report credible performance evaluation results
  obtained for 204600 randomly-generated passive optical networks.  We
  show that with the proposed sharing, operators can increase their
  access network performance even \emph{twofold} with software-defined
  upgrades, and with no or minimal new hardware required.
\end{abstract}


\section{Introduction}


New network architectures and business models are required to further
increase the network performance and availability, and still to cut
costs and save energy.  One of them is sharing, being both a network
architecture and a business model \cite{10.1109/JPROC.2014.2302743}.
Currently, network operators do share, but the physical infrastructure
only (buildings, towers, etc.) \cite{10.1109/MCOM.2011.6035827}.  Some
operators \emph{merge} their networks into a single network to own and
use it in a marriage-like fashion.

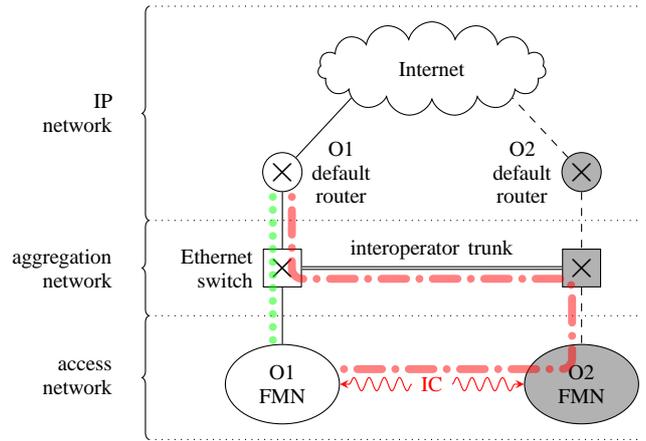
\begin{figure}
  \centering
  \begin{tikzpicture}
  [node distance = 0.75 cm and 0.75 cm, font = \footnotesize]

  \node [mycloud] (internet) at (0, 0)
  {Internet};
  \node [router, font = \Large] (r1)
  [below left = of internet]
  {$\times$};
  \node [router, font = \Large, fill = black!30] (r2)
  [below right = of internet]
  {$\times$};

  \draw (r1) -- (internet.200);
  \draw [dashed] (r2) -- (internet.-20);

  \node [align = center] (r1i) [right = 0 cm of r1]
  {O1\\default\\router};
  \node [align = center] (r2i) [left = 0 cm of r2]
  {O2\\default\\router};

  \node [switch, font = \Large] (s1)
  [below = of r1]
  {$\times$};
  \node [switch, font = \Large, fill = black!30] (s2)
  [below = of r2]
  {$\times$};

  \node [align = right] (es) [left = 0 cm of s1]
  {Ethernet\\switch};

  \draw (r1) -- (s1);
  \draw [dashed] (r2) -- (s2);
  \draw [double] (s1) -- node [above] {interoperator trunk} (s2);

  \node [fmn] (fmn1) [below = of s1] {O1\\FMN};
  \node [fmn, fill = black!30] (fmn2) [below = of s2] {O2\\FMN};

  \draw (s1) -- (fmn1);
  \draw [dashed] (s2) -- (fmn2);

  \node [red] (ic) at ($ (fmn1)!0.5!(fmn2) $) {IC};
  \draw [icwave] (ic) -- (fmn1);
  \draw [icwave] (ic) -- (fmn2);

  \coordinate (lbr) at ($ (fmn1.west) - (1 cm, 0) $);
  \coordinate (rbr) at (fmn2.east);

  \coordinate (cla) at ($ (internet.north) + (0, 0.2 cm) $);
  \coordinate (cla1) at (cla -| lbr);
  \coordinate (cla2) at (cla -| rbr);
  \draw [dotted] (cla1) -- (cla2);

  \coordinate (clb) at ($ (r1.south)!0.5!(s1.north) $);
  \coordinate (clb1) at (clb -| lbr);
  \coordinate (clb2) at (clb -| rbr);
  \draw [dotted] (clb1) -- (clb2);

  \coordinate (clc) at ($ (s1.south)!0.5!(fmn1.north) $);
  \coordinate (clc1) at (clc -| lbr);
  \coordinate (clc2) at (clc -| rbr);
  \draw [dotted] (clc1) -- (clc2);

  \coordinate (cld) at ($ (fmn1.south) - (0, 0.2 cm) $);
  \coordinate (cld1) at (cld -| lbr);
  \coordinate (cld2) at (cld -| rbr);
  \draw [dotted] (cld1) -- (cld2);

  \draw [decorate, decoration = brace]
  (clb1) -- node [left = 10 pt, align = right]
  {IP\\network} (cla1);

  \draw [decorate, decoration = brace]
  (clc1) -- node [left = 10 pt, align = right]
  {aggregation\\network} (clb1);

  \draw [decorate, decoration = brace]
  (cld1) -- node [left = 10 pt, align = right]
  {access\\network} (clc1);

  \def\dfnca{0.125 cm}
  \def\dfncb{0.2 cm}

  \draw [dash pattern = on 0pt off 5pt on 9.225pt off 5pt, line width =
  3 pt, red, opacity = 0.5, rounded corners = 4 pt, line cap =
  round]
  ($ (fmn1.east) + (1.5 pt, \dfncb) $) --
  ($ (fmn2) + (-\dfnca, \dfncb) $) --
  ($ (s2) + (-\dfnca, -\dfnca - 0.5 pt) $) --
  ($ (s1) + (\dfnca, -\dfnca - 0.5 pt) $) --
  ($ (r1.south) + (\dfnca, -1.5 pt) $);

  \draw [dash pattern = on 0 off 4.9 pt, line width = 3 pt, green,
  opacity = 0.5, rounded corners = 4 pt, line cap = round]
  ($ (fmn1.north) + (-\dfnca, 1.5 pt) $) --
  ($ (r1.south) + (-\dfnca, -1.5 pt) $);

\end{tikzpicture}
  \caption[General network architecture]{General network architecture.}
  \label{f:general}
\end{figure}

We proposed the interoperator fixed-mobile network (FMN) sharing, and
demonstrated impressive \emph{availability} improvement
\cite{ondm2015}.  Fig.~\ref{f:general} illustrates the proposed
sharing, where one operator (O1) can divert its access-network traffic
to the other operator (O2) with the \emph{interoperator communication}
(IC), either for performance or availability reasons.  The diverted
traffic is then sent back to O1 through the aggregation network with
either Q-in-Q or MAC-in-MAC tagging.

While the proposed sharing could be used with any access network
technology, we concentrate on the passive optical networks (PONs),
because they are economical and wide-spread.  PONs are also attractive
for the emerging fifth-generation (5G) radio access networks
\cite{10.1109/SURV.2013.013013.00135}.  The sharing can be implemented
by the interoperator-communicating optical network units (IC-ONUs),
i.e., the interconnected ONUs of different operators.  An IC-ONU can
offer the communication with the aggregation network in the same way
as the optical line terminal (OLT) does, while the remaining ONUs are
non interoperator-communicating ONUs (NIC-ONUs).  The data of the
first operator can be diverted by an active remote node (RN) to the
access network of the other operator.

\begin{figure}
  \centering
  \begin{tikzpicture}[fill = black, scale = 0.9]

  \node [circ1] (co1) at (0, 0) {CO};
  \node [circ2] (co2) at (7.5, 0) {CO};

  \node [prn] (cs1a) at ($ (co1) + (1.25, 0.25) $) {};
  \node [arn] (cs1b) at ($ (co1) + (4, 0.5) $) {};
  \node [prn] (cs2a) at ($ (co2) - (1, 0) $) {};
  \node [arn] (cs2b) at ($ (co2) - (4, 0.5) $) {};

  \node [circ1] (bs1a) at (0.75, 2) {\BS{scale = 0.1}};
  \node [circ1] (bs1b) at (7, 2) {\BS{scale = 0.1}};
  \node [circ1] (bs1d) at (1.5, -2) {\BS{scale = 0.1}};
  \node [circ1] (ho1a) at (2, -0.5) {\FU{scale = 0.09}};
  \node [circ1] (ho1b) at (3.75, 1.25) {\FU{scale = 0.09}};
  \node [circ1] (ho1c) at (4.5, -1.5) {\FU{scale = 0.09}};

  \node [circ2] (bs2b) at (4.5, 2) {\BS{scale = 0.1}};
  \node [circ2] (bs2c) at (0.75, -1.5) {\BS{scale = 0.1}};
  \node [circ2] (ho2a) at (3.25, -1.75) {\FU{scale = 0.09}};
  \node [circ2] (ho2b) at (2.5, 1) {\FU{scale = 0.09}};
  \node [circ2] (ho2c) at (7, -1.5) {\FU{scale = 0.09}};

  \node [circle, inner sep = -0.2 mm] (mu1)
  at (6.5, 0.5) {\MU{scale = 0.15}};
  \node [circle, inner sep = -0.2 mm] (mu2)
  at (1, -0.25) {\MU{scale = 0.15}};
  
  \draw (co1) -- (cs1a) -- (cs1b);
  \draw (cs1a) -- (bs1a);
  \draw (cs1a) -- (bs1d);
  \draw (cs1a) -- (ho1a);
  \draw (cs1b) -- (bs1b);
  \draw (cs1b) -- (ho1b);
  \draw (cs1b) -- (ho1c);

  \draw[dashed] (co2) -- (cs2a) -- (cs2b);
  \draw [dashed] (cs2a) -- (ho2c);
  \draw [dashed] (cs2a) -- (bs2b);
  \draw [dashed] (cs2b) -- (ho2a);
  \draw [dashed] (cs2b) -- (bs2c);
  \draw [dashed] (cs2b) -- (ho2b);

  \node [red] (ic1) at ($ (bs1b)!0.5!(bs2b) $) {IC};
  \draw [icwave] (ic1) -- (bs1b);
  \draw [icwave] (ic1) -- (bs2b);
  \node [red] (ic2) at ($ (bs1a)!0.5!(bs2b) $) {IC};
  \draw [icwave] (ic2) -- (bs1a);
  \draw [icwave] (ic2) -- (bs2b);

  \node [red] (ic3) at ($ (ho1c)!0.5!(ho2c) $) {IC};
  \draw [iclink] (ic3) -- (ho1c);
  \draw [iclink] (ic3) -- (ho2c);

  \draw [muwave] (mu1) -- (bs1b);
  \draw [muwave] (mu2) -- (bs2c);

\end{tikzpicture}
  \caption[Interoperator FMN sharing]{Interoperator FMN sharing: a
    central office (CO), the interoperator communication (IC), a
    passive remote node $\bullet$, an active remote node $\circ$, a
    mobile user \MU{scale = 0.12}, a base station \BS{scale = 0.05},
    and a fixed user \FU{scale = 0.05}.}
  \label{f:sharing}
\end{figure}
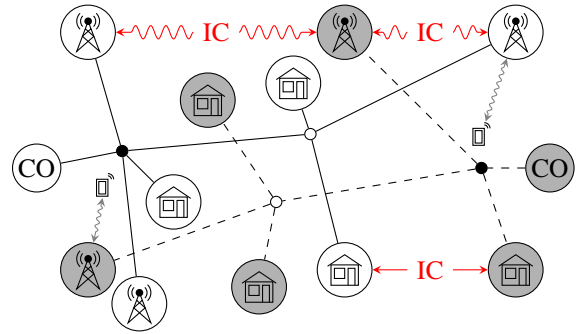


\emph{The contribution of our work is the evaluation of the
performance improvement brought by our sharing along with the
evaluation software under the General Public License
\cite{perforwebsite}.}

In Section \ref{related} we review related works, in Section
\ref{problem} we state the problem, in Section \ref{algorithm} we
propose a solution to the problem, which we harness in Section
\ref{evaluation} to produce credible performance results.  Finally,
Section \ref{conclusion} concludes the article.


\section{Related works}
\label{related}


Mobile network sharing has long been used, allowing for roaming or
virtual mobile network operators, where a mobile operator accepts
traffic \emph{directly from the users} of a different operator, and
the network of that different operator does not carry that traffic.
In \cite{10.1109/MCOM.2011.6035827} the authors study this traditional
sharing in the virtualization context.  Our work does not subscribe to
this mainstream research -- the hallmark of our proposed sharing is
the \emph{interoperator communication}, where traffic is exchanged by
different operators between their access networks.


PONs are successful mainly because of the cost-effective tree
topology.  First, the feeder fiber starts at the optical line terminal
(OLT) in the central office (CO), and ends at the first remote node
(RN) in some district.  From there, the distribution fibers lead to
further RNs in various neighborhoods, possibly through further RNs.
Finally, the last-mile fibers deliver the service to customer
premises.


NG-PONs should support direct communication between ONUs, without the
OLT relaying the data, in order to support direct communication
between BSs (connected to ONUs) required by future RANs.  However, in
legacy PONs, ONUs do not communicate directly with each other, but
through the OLT.  To this end, in \cite{10.1109/JLT.2010.2050861} the
authors propose two novel NG-PON architectures.  Interestingly, the
authors propose to cleverly use a circulator as a passive RN, which
would allow for some limited communication between BSs without the
OLT.  Another solution is to use the active RNs, which would also
enable NG-PONs to have larger splitting ratios and longer reach
\cite{10.1364/JOCN.2.000028}.


In \cite{10.1109/TR.2011.2134210}, the authors propose a number of
wireless protection methods for FMNs, which could also be used to
increase performance.  There a single network is considered, without
sharing it with a different operator.  Wireless access points
connected to a PON are allowed to offer connectivity to those wireless
access points which lost the PON connectivity.


PONs are vulnerable to service disruption, because of the tree
architecture.  Failure of the OLT or the feeder fiber brings down the
entire PON.  Making a PON resilient is becoming more important, but
requires expensive redundant infrastructure, fibers and hardware.  In
\cite{10.1109/MCOM.2014.6736740} the authors review PON resiliency
mechanisms and propose their own mechanism for cost-effective
resiliency on request.  \emph{The redundant hardware can be used to
improve the PON performance.}


FMNs are being broadly researched and developed to deliver the
required performance and resiliency \cite{10.1109/JPROC.2012.2185769}.
Radio access technologies (RATs) have been proposed to use
cognitivity, virtualization, coordinated multipoint transmission
(CoMT), and more sophisticated modulation formats.  The backhaul is
evolving from the copper or microwave networks to passive optical
networks (PONs), and even possibly to radio-over-fiber (RoF) networks.
PONs are currently being deployed as the backhaul, and the NG-PONs are
being intensively researched for FMNs \cite{10.1109/JLT.2010.2050861}.


\section{Problem statement}
\label{problem}


We are given a specific PON $P$, i.e., a PON with the topology, and
the types of the RNs and ONUs given.  The number of ONUs is $N$, the
number of RNs is $R$, and the downstream capacity is $c$.


The problem is to evaluate the \emph{upper bound} of the downstream
performance $p$ of PON $P$ under the offered load $1 \le l \le 2$,
where $l = 1$ is the full offered load, and $l = 2$ is twice the full
offered load.  We define $p$ as the average of the downstream
performance values of all ONUs in a PON.  In turn, we define the
downstream ONU performance as the ratio of the downstream bitrate (DB)
granted to an ONU to the DB requested by that ONU.  The DB can be
granted either by the OLT or by an IC-ONU.  When $P$ is able to
completely service the offered load $l$, performance $p = 1$.  Even
though the proposed evaluation algorithm (described in Section
\ref{algorithm}) can handle non-uniform traffic, for simplicity, we
load the PON with uniform traffic: we try to grant the same DB $b =
cl/N$ to every ONU.


The DB granted by the OLT is guaranteed to reach the aggregation
network.  However, we assume that the DB granted by an IC-ONU is
\emph{potential} only: even though the bitrate is granted by an IC-ONU
within the PON, the other operator might actually reject it.  The
\emph{sharing rules} should regulate whether the potential DB is
guaranteed by the other operator.  We make this assumption to study
the upper bound of the performance, without considering the sharing
rules.

\begin{figure}
  \centering
  \begin{tikzpicture}[fill = black]

  \node [circle] (co) at (0, 0) {\VDOTS{rotate = 90}};

  \node [arn] (rn1) at ($ (co) + (0.9, 0) $) {};
  \node [prn] (rn2) at ($ (co) + (2.7, 0) $) {};
  \node [prn] (rn3) at ($ (co) + (4.5, 0) $) {};

  \node [font = \footnotesize] at ($ (rn1) - (0.2 cm, 0.35 cm) $) {RN1};
  \node [font = \footnotesize] at ($ (rn2) - (0.2 cm, 0.35 cm) $) {RN2};
  \node [font = \footnotesize] at ($ (rn3) - (0.2 cm, 0.35 cm) $) {RN3};

  \def\d2r{1 cm}

  \draw (co) -- (rn1);
  \node [circle] (r1a) at ($ (rn1) + (0.75 cm, 0.75 cm) $) {\VDOTS{rotate = -45}};
  \node [circle] (r1b) at ($ (rn1) + (0.75 cm, -0.75 cm) $) {\VDOTS{rotate = 45}};
  \draw (rn1) -- (r1a);
  \draw (rn1) -- (rn2);
  \draw (rn1) -- (r1b);

  \node [circ1, label = IC-ONU] (r2a) at ($ (rn2) + (\d2r, \d2r) $) {\BS{scale = 0.1}};
  \node [circle] (r2b) at ($ (rn2) + (0.75 cm, -0.75 cm) $) {\VDOTS{rotate = 45}};
  \draw (rn2) -- (r2a);
  \draw (rn2) -- (rn3);
  \draw (rn2) -- (r2b);

  \node [circ1, label = NIC-ONU] (r3a) at ($ (rn3) + (\d2r, \d2r) $) {\FU{scale = 0.09}};
  \node [circle] (r3b) at ($ (rn3) + (0.75 cm, -0.75 cm) $) {\VDOTS{rotate = 45}};
  \draw (rn3) -- (r3a);
  \draw (rn3) -- (r3b);

  \def\dfnca{0.15 cm}

  \path [name path = p1a]
  ($ (rn2) + (-45:\dfnca) $) -- ($ (r2a.-135) + (-45:\dfnca) $);

  \path [name path = p1b]
  ($ (rn2) + (90:\dfnca) $) -- ($ (rn3) + (90:\dfnca) $);

  \path [name intersections = {of = p1a and p1b}];
  \coordinate (c1a) at (intersection-1);

  \path [name path = p1c]
  ($ (rn3) + (135:\dfnca) $) -- ($ (r3a.-135) + (135:\dfnca) $);

  \path [name intersections = {of = p1b and p1c}];
  \coordinate (c1b) at (intersection-1);

  \draw [black, dash pattern = on 6 pt off 5.5 pt, line width = 3 pt,
    opacity = 0.5, rounded corners = 4 pt, line cap = round]
  ($ (r2a.-135) + (-45:\dfnca) + (-135:\dfnca) $) --
  (c1a) -- (c1b) --
  ($ (r3a.-135) + (135:\dfnca) + (-135:\dfnca) $);

  \draw [red, dash pattern = on 6 pt off 5.5 pt, line width = 2.5 pt,
    opacity = 0.5, rounded corners = 4 pt, line cap = round]
  ($ (r2a.-135) + (-45:\dfnca) + (-135:\dfnca) $) --
  (c1a) -- (c1b) --
  ($ (r3a.-135) + (135:\dfnca) + (-135:\dfnca) $);

  \path [name path = p2a]
  ($ (rn2) + (135:\dfnca) $) -- ($ (r2a.-135) + (135:\dfnca) $);

  \path [name path = p2b]
  ($ (rn1) + (90:\dfnca) $) -- ($ (rn2) + (90:\dfnca) $);

  \path [name intersections = {of = p2a and p2b}];
  \coordinate (c2a) at (intersection-1);

  \path [name path = p2c]
  ($ (rn1) + (-45:\dfnca) $) -- ($ (r1a) + (-45:\dfnca) $);

  \path [name intersections = {of = p2b and p2c}];
  \coordinate (c2b) at (intersection-1);

  \path [name path = p2d]
  ($ (rn1) + (-90:\dfnca) $) -- ($ (rn3) + (\dfnca, -\dfnca) $);

  \path [name path = p2e]
  ($ (rn1) + (45:\dfnca) $) -- ($ (r1b) + (45:\dfnca) $);

  \path [name intersections = {of = p2d and p2e}];
  \coordinate (c2c) at (intersection-1);

  \path [name path = p2f]
  ($ (rn3) + (-45:\dfnca) - (\dfnca, \dfnca) $) -- ($ (r3a.-135) + (-45:\dfnca) $);

  \path [name intersections = {of = p2d and p2f}];
  \coordinate (c2d) at (intersection-1);

  \draw [black, dash pattern = on 0 pt off 5 pt on 10 pt off 5 pt,
    line width = 3 pt, opacity = 0.5, rounded corners = 4 pt, line
    cap = round]
  ($ (r2a.-135) + (135:\dfnca) + (-135:\dfnca) $) --
  (c2a) -- (c2b) -- (c2c) -- (c2d) --
  ($ (r3a.-135) + (-45:\dfnca) + (-135:\dfnca) $);

  \draw [green, dash pattern = on 0 pt off 5 pt on 10 pt off 5 pt,
    line width = 2.5 pt, opacity = 0.5, rounded corners = 4 pt, line
    cap = round]
  ($ (r2a.-135) + (135:\dfnca) + (-135:\dfnca) $) --
  (c2a) -- (c2b) -- (c2c) -- (c2d) --
  ($ (r3a.-135) + (-45:\dfnca) + (-135:\dfnca) $);

\end{tikzpicture}
  \caption[Shortest path example]{Example of finding a correct
    shortest path between an IC-ONU and an NIC-ONU, where $\bullet$ is
    a passive RN, $\circ$ is an active RN, \raisebox{1.5pt}{\RP} is
    the correct path, and \raisebox{1.5pt}{\WP} is the wrong path.}
  \label{f:example}
\end{figure}
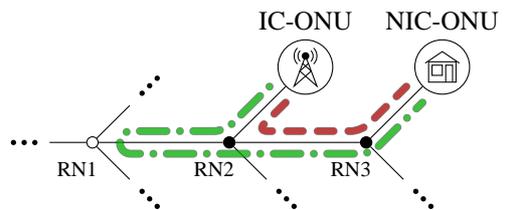

Fig.~\ref{f:example} illustrates the corollary problem of finding a
correct shortest path from an IC-ONU to an NIC-ONU traversing a
passive RN.  It is a problem unusual for legacy PONs where downstream
data frames travel downstream only, while with the proposed sharing
the downstream data frames sent from an IC-ONU travel upstream first,
are diverted by an active RN, and then travel downstream to finally
reach an NIC-ONU.  If we apply a shortest path algorithm (e.g., the
breadth-first search algorithm or the Dijkstra algorithm) directly on
$P$, we get the dashed (red) path, which is wrong, instead of the
dash-dotted (green) path, which is correct.

In the next section we describe the algorithm which evaluates the
performance of a single PON.  To make the evaluation of the proposed
sharing statistically sound, in Section \ref{evaluation} we use the
algorithm to evaluate the performance of PON populations.


\section{Algorithm}
\label{algorithm}


In order to evaluate performance $p$, we need to calculate performance
$p_n$ of ONU $n$.  In a PON without sharing, the DB is granted by the
OLT only, but with sharing, an ONU can have the DB granted also by
IC-ONUs, and so the performance evaluation is more complicated, as we
have to consider all the alternatives an ONU has.  An
\emph{alternative} for an ONU is a shortest path (i.e., a path with
the smallest number of hops) to that ONU from either the OLT or an
IC-ONU.  An ONU has at least one alternative: the one from the OLT.


\begin{algorithm}[t]
  \caption{calculate performance\\
    \textbf{Input:} PON $P$, load $l$, capacity $c$, number of ONUs $N$\\
    \textbf{Output:} performance $p$}
  \label{a:algorithm}
  \begin{algorithmic}
    \State $\textbf{A} = (A_n) \gets \text{find alternatives }(P)$
    \State $Q \gets \text{make priority queue $(\textbf{A})$}$
    \State $b \gets cl / N$
    \ForEach{$n \in \lbrace\text{ONUs of $P$}\rbrace$}
    \State $Q.\text{push }(n)$
    \EndFor
    \While {$Q \ne \emptyset$}
    \State $n \gets Q.\text{pop }()$
    \State $p_n \gets \text{grant DB }(P, n, b, A_n)/b$
    \EndWhile
    \State\Return $p \gets \text{average }(p_n)$
  \end{algorithmic}
\end{algorithm}


PON performance can be evaluated in various ways and with different
accuracy depending on the PON type, the traffic type, and a plethora
of technical details.  We propose a simple method of assessing the
performance: we start with an unloaded PON, and try to grant the same
DB $b$ to every ONU.


We serve the ONUs in the order of increasing number of alternatives
they have, since the more alternatives an ONU has, the more likely it
is to get the DB.  If an ONU being served has more than one
alternative, we try to grant the DB along the shortest alternative,
and continue with the longer alternatives when needed.  The DB is
granted to an ONU by allocating the DB on the edges of a single
alternative.


Algorithm \ref{a:algorithm} recaps the described performance
evaluation.  Function 'find alternatives' returns vector $\textbf{A}$
of alternatives $A_n$ for ONU $n$ in $P$ sorted in the increasing
order of the number of hops.  $Q$ is the priority queue of ONUs sorted
in the increasing order of the number of alternatives the ONUs have.
$Q$ is given the alternatives $\textbf{A}$ at the construction time,
and then uses $\textbf{A}$ to maintain the order of the nodes pushed
into $Q$.  Function 'grant DB' tries to grant DB $b$ to ONU $n$ (i.e.,
it allocates the bandwidth on all the edges of an alternative), and
returns the value of the actually granted DB.  The 'while' loop
iterates over the ONUs requesting service, finishing when there are no
more ONUs to service.


In the search for the correct alternatives, we use the breadth-first
search algorithm (BFS) on a modified graph as described next.
However, a different solution would be to adapt the BFS algorithm:
make the label of a passive RN also take the direction with which that
passive RN was reached.  However, we opt for modifying $P$, since we
find it more elegant.

\begin{figure}
  \centering
  \begin{tikzpicture}[fill = black, > = stealth]

  \node [circle] (co) at (0, 0) {\VDOTS{rotate = 90}};

  \node [arn] (rn1) at ($ (co) + (0.9, 0) $) {};

  \coordinate (rn2) at ($ (co) + (2.7, 0) $);
  \node [sprn] (rn2u) at ($ (rn2) + (0, 0.15) $) {};
  \node [sprn] (rn2d) at ($ (rn2) + (0, -0.15) $) {};

  \coordinate (rn3) at ($ (co) + (4.5, 0) $);
  \node [sprn] (rn3u) at ($ (rn3) + (0, 0.15) $) {};
  \node [sprn] (rn3d) at ($ (rn3) + (0, -0.15) $) {};

  \node [font = \footnotesize] at ($ (rn1) - (0.2 cm, 0.35 cm) $) {RN1};
  \node [font = \footnotesize] at ($ (rn2) - (0.2 cm, 0.35 cm) $) {RN2};
  \node [font = \footnotesize] at ($ (rn3) - (0.2 cm, 0.35 cm) $) {RN3};

  \def\d2r{1 cm}

  \draw (co) -- (rn1);
  \node [circle] (r1a) at ($ (rn1) + (0.75 cm, 0.75 cm) $) {\VDOTS{rotate = -45}};
  \node [circle] (r1b) at ($ (rn1) + (0.75 cm, -0.75 cm) $) {\VDOTS{rotate = 45}};
  \draw (rn1) -- (r1a);
  \draw (rn1) edge[<-, bend left = 5] (rn2u);
  \draw (rn2d) edge[<-, bend left = 5] (rn1);
  \draw (rn1) -- (r1b);

  \node [circ1, label = IC-ONU] (r2a) at ($ (rn2) + (\d2r, \d2r) $) {\BS{scale = 0.1}};
  \node [circle] (r2b) at ($ (rn2) + (0.75 cm, -0.75 cm) $) {\VDOTS{rotate = 45}};

  \draw (rn2u) edge[<-, bend left = 5] (r2a);
  \draw (r2a) edge[<-, bend left = 5] (rn2d);

  \draw [<-] (rn2u) -- (rn3u);
  \draw [<-] (rn3d) -- (rn2d);

  \draw (rn2u) edge[<-, bend left = 5] (r2b);
  \draw (r2b) edge[<-, bend left = 5] (rn2d);

  \node [circ1, label = NIC-ONU] (r3a) at ($ (rn3) + (\d2r, \d2r) $) {\FU{scale = 0.09}};
  \node [circle] (r3b) at ($ (rn3) + (0.75 cm, -0.75 cm) $) {\VDOTS{rotate = 45}};

  \draw (rn3u) edge[<-, bend left = 5] (r3a);
  \draw (r3a) edge[<-, bend left = 5] (rn3d);

  \draw (rn3u) edge[<-, bend left = 5] (r3b);
  \draw (r3b) edge[<-, bend left = 5] (rn3d);

  \def\dfnca{0.15 cm}

  \path [name path = p2a]
  ($ (rn2) + (135:\dfnca) $) -- ($ (r2a.-135) + (135:\dfnca) $);

  \path [name path = p2b]
  ($ (rn1) + (90:\dfnca) $) -- ($ (rn2) + (90:\dfnca) $);

  \path [name intersections = {of = p2a and p2b}];
  \coordinate (c2a) at (intersection-1);

  \path [name path = p2c]
  ($ (rn1) + (-45:\dfnca) $) -- ($ (r1a) + (-45:\dfnca) $);

  \path [name intersections = {of = p2b and p2c}];
  \coordinate (c2b) at (intersection-1);

  \path [name path = p2d]
  ($ (rn1) + (-90:\dfnca) $) -- ($ (rn3) + (\dfnca, -\dfnca) $);

  \path [name path = p2e]
  ($ (rn1) + (45:\dfnca) $) -- ($ (r1b) + (45:\dfnca) $);

  \path [name intersections = {of = p2d and p2e}];
  \coordinate (c2c) at (intersection-1);

  \path [name path = p2f]
  ($ (rn3) + (-45:\dfnca) - (\dfnca, \dfnca) $) -- ($ (r3a.-135) + (-45:\dfnca) $);

  \path [name intersections = {of = p2d and p2f}];
  \coordinate (c2d) at (intersection-1);

  \draw [black, dash pattern = on 0 pt off 5 pt on 10 pt off 5 pt,
    line width = 3 pt, opacity = 0.25, rounded corners = 4 pt, line
    cap = round]
  ($ (r2a.-135) + (135:\dfnca) + (-135:\dfnca) $) --
  (c2a) -- (c2b) -- (c2c) -- (c2d) --
  ($ (r3a.-135) + (-45:\dfnca) + (-135:\dfnca) $);

  \draw [green, dash pattern = on 0 pt off 5 pt on 10 pt off 5 pt,
    line width = 2.5 pt, opacity = 0.25, rounded corners = 4 pt, line
    cap = round]
  ($ (r2a.-135) + (135:\dfnca) + (-135:\dfnca) $) --
  (c2a) -- (c2b) -- (c2c) -- (c2d) --
  ($ (r3a.-135) + (-45:\dfnca) + (-135:\dfnca) $);

\end{tikzpicture}
  \caption[Splitting example]{Example of modifying a graph by
    splitting passive RNs, where \raisebox{1.5pt}{\sprn} is a split
    passive RN, $\circ$ is an active RN, and \raisebox{1.5pt}{\RP} is
    the correct path.}
  \label{f:splitting}
\end{figure}
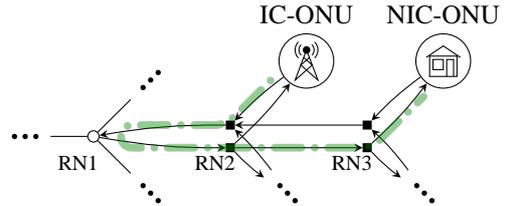

We find the correct alternatives by searching for shortest paths in a
\emph{modified} graph $P'$, where a passive RN is modeled with two
vertexes: one vertex for the downstream direction with the
downstream-directed edges only, and the other vertex for the upstream
direction with the upstream-directed edges only, as shown in
Fig.~\ref{f:splitting}.  A shortest-path search from a given IC-ONU
keeps going upstream through a passive RN (without being derailed
there) to an active RN, where it turns back to continue downstream,
passes again through the passive RN, and finally reaches an NIC-ONU.

\begin{algorithm}[t]
  \caption{find alternatives\\
    \textbf{Input:} PON $P$\\
    \textbf{Output:} alternatives $\textbf{A}$}
  \label{a:alternatives}
  \begin{algorithmic}
    \State $\textbf{A} = (A_n \gets \emptyset)$
    \State $P' \gets \text{split RNs}(P)$
    \ForEach{$i \in \lbrace\text{OLT and IC-ONUs of $P$}\rbrace$}
    \State $T_i \gets \text{breadth-first search }(P', i)$
    \ForEach{$n \in \lbrace\text{ONUs of $P$}\rbrace$}
    \State $A_{i, n} \gets \text{trace } (T_i, n)$
    \If {$A_{i, n}$ exists}
    \State $A_n.\text{push } (A_{i, n})$
    \EndIf
    \EndFor
    \EndFor
    \State\Return $\textbf{A} = (A_n)$
  \end{algorithmic}
\end{algorithm}

For every node $i$ providing service, which is the OLT or an IC-ONU,
we search for the shortest-path tree $T_i$ from node $i$ to all other
nodes of the PON, where the search stops.  The search starting at an
IC-ONU also stops at the OLT, even though the OLT could divert
downstream the upstream traffic to reach an NIC-ONU, since the OLT is
an active node.  This, however, would be pointless, since the OLT can
service the NIC-ONU, without the IC-ONU, and without taking the
upstream capacity.

Algorithm \ref{a:alternatives} calculates the alternatives, returned
as vector $\textbf{A}$.  $A_n$ is a priority queue, where alternatives
for node $n$ are sorted in the increasing order of their number of
hops.  There are two nested loops.  The outer loop iterates for each
node $i$ providing service, and the inner loop iterates for each node
$n$ receiving service.  In the outer loop the shortest-path tree $T_i$
is calculated, which is subsequently used in the inner loop to find,
using function 'trace', a single shortest path $A_{i, n}$ from node
$i$ to node $n$.  We make sure $A_{i, n}$ does exist, since it could
very well be that it does not, as when there is no active RN on the
way from an IC-ONU to an NIC-ONU.


\section{Performance evaluation}
\label{evaluation}


We evaluate the performance of a given PON population under a given
offered load $l$.  In order to obtain credible estimated results for a
population, we produce a sample of 300 specific randomly-generated
PONs.  Performance $p$ is calculated by the proposed algorithm for
every PON in the sample under load $l$, and then the obtained PON
performance values are averaged to yield the sample PON performance
mean.  We deem the obtained means credible, as their relative standard
errors are below 1\%.

\begin{figure}
  \centering
  \begin{tikzpicture}[fill = black, scale=0.95]

  \node [circ1] (co) at (0, 0) {CO};

  \node [prn] (rn1) at ($ (co) + (1.05, 0) $) {};
  \node [arn] (rn2) at ($ (co) + (3.3, 0) $) {};
  \node [prn] (rn3) at ($ (co) + (5.55, 0) $) {};

  \node [font = \footnotesize] at ($ (rn1) - (0.2 cm, 0.3 cm) $) {1:g};
  \node [font = \footnotesize] at ($ (rn2) - (0.2 cm, 0.3 cm) $) {1:g};
  \node [font = \footnotesize] at ($ (rn3) - (0.2 cm, 0.3 cm) $) {1:g};

  \def\d2r{1.25 cm}

  \node [circ1] (r1a) at ($ (rn1) + (\d2r, 1.25 cm) $) {\BS{scale = 0.1}};
  \node [circ1] (r1b) at ($ (rn1) + (\d2r, -1.25 cm) $) {\FU{scale = 0.09}};
  \node at ($ (r1a.south)!0.25!(r1b.north) $) {\VDOTS{}};
  \node at ($ (r1a.south)!0.75!(r1b.north) $) {\VDOTS{}};

  \node [circ1] (r2a) at ($ (rn2) + (\d2r, 1.25 cm) $) {\FU{scale = 0.09}};
  \node [circ1] (r2b) at ($ (rn2) + (\d2r, -1.25 cm) $) {\BS{scale = 0.1}};
  \node at ($ (r2a.south)!0.25!(r2b.north) $) {\VDOTS{}};
  \node at ($ (r2a.south)!0.75!(r2b.north) $) {\VDOTS{}};

  \node [circ1] (r3a) at ($ (rn3) + (\d2r, 1.25 cm) $) {\BS{scale = 0.1}};
  \node [circ1] (r3b) at ($ (rn3) + (\d2r, 0.25 cm) $) {\FU{scale = 0.09}};
  \node [circ1] (r3c) at ($ (rn3) + (\d2r, -1.25 cm) $) {\FU{scale = 0.09}};
  \node at ($ (r3b)!0.5!(r3c) $) {\VDOTS{}};

  \draw (co) -- (rn1);
  \node [font = \footnotesize] (s1a)
  at ($ (rn1)!0.5!(r1a) $) {$1 - s$};
  \draw (rn1) -- (s1a) -- (r1a);
  \node [font = \footnotesize] (s1b)
  at ($ (rn1)!0.5!(rn2) $) {$s$};
  \draw (rn1) -- (s1b) -- (rn2);
  \node [font = \footnotesize] (s1c)
  at ($ (rn1)!0.5!(r1b) $) {$1 - s$};
  \draw (rn1) -- (s1c) -- (r1b);

  \node [font = \footnotesize] (s2a)
  at ($ (rn2)!0.5!(r2a) $) {$1 - s$};
  \draw (rn2) -- (s2a) -- (r2a);
  \node [font = \footnotesize] (s2b)
  at ($ (rn2)!0.5!(rn3) $) {$s$};
  \draw (rn2) -- (s2b) -- (rn3);
  \node [font = \footnotesize] (s2c)
  at ($ (rn2)!0.5!(r2b) $) {$1 - s$};
  \draw (rn2) -- (s2c) -- (r2b);

  \draw (rn3) -- (r3a);
  \draw (rn3) -- (r3b);
  \draw (rn3) -- (r3c);

  \coordinate (d1) at ($ (rn1) - (0.5 cm, 1.75 cm) $);
  \coordinate (d2) at ($ (rn2) - (0.5 cm, 1.75 cm) $);
  \coordinate (d3) at ($ (rn3) - (0.5 cm, 1.75 cm) $);
  \coordinate (d4) at ($ (rn3) - (-1.75 cm, 1.75 cm) $);
  \draw [dotted] (d1) -++ (0, 3.5 cm);
  \draw [dotted] (d2) -++ (0, 3.5 cm);
  \draw [dotted] (d3) -++ (0, 3.5 cm);
  \draw [dotted] (d4) -++ (0, 3.5 cm);
  \draw [decorate, decoration = brace]
  (d2) -- node [below = 5 pt, align = center]
        {\nth{1} stage} (d1);
        \draw [decorate, decoration = brace]
        (d3) -- node [below = 5 pt, align = center]
              {\nth{2} stage} (d2);
              \draw [decorate, decoration = brace]
              (d4) -- node [below = 5 pt, align = center]
                    {\nth{3} stage} (d3);

\end{tikzpicture}
  \caption[PON]{PON: a central office
  (CO), a passive remote node $\bullet$, an active remote node
  $\circ$, a base station \BS{scale = 0.05}, and a fixed user
  \FU{scale = 0.05}.}
  \label{f:scenario}
\end{figure}
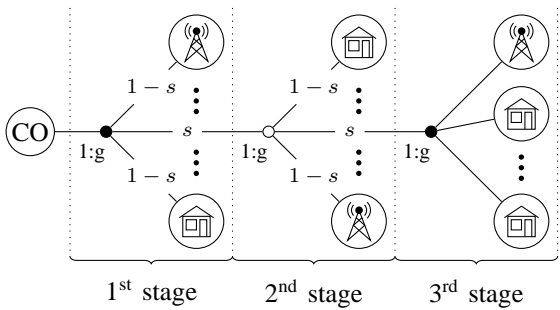

We evaluate the performance improvement when the proposed sharing is
used for PON populations of two scenarios.  Both scenarios use
populations of a generic PON shown in Fig.~\ref{f:scenario} with the
depth of three stages.  The PON can have many second-stage and
third-stage RNs, but in the figure we show only one of each.  The
scenarios differ in the location of the active RNs: in the first
scenario the location is given upfront, while in the second scenario
the location is chosen at random.


The RNs of every stage have the \mbox{1:g} splitting ratio.  At the
first and the second stage, a fiber coming out of a RN goes to the
next stage with probability $s$, and, conversely, to an ONU, i.e., a
fixed user or a base station, with probability $(1 - s)$.  At the
third stage, all fibers coming out of a RN go to ONUs.

For the studied PON network, the mean number of ONUs is $N = g(1 - s +
gs(1 - s + gs))$, and $R = 1 + gs(1 + gs)$.  For the evaluation we
assume $s = 0.3$ and $g = 32$, and so $N \approx 3187$ and $R \approx
103$, which is reasonable for the next-generation PONs.  For example,
the XG-PON currently can handle 1024 ONUs.  Furthermore, we assumed
the PON offers the downstream capacity of $c = 10$ Gb/s, and the
upstream capacity of 2.5 Gb/s.  IC-ONUs can send data to the other
operator with the bitrate of 2.5 Gb/s.


In Figures \ref{f:1st} and \ref{f:2nd}, we plot as the white surface
the PON performance with the proposed sharing, and as the gray surface
the PON performance without the proposed sharing.  The difference
between these surfaces is the performance improvement brought by our
sharing.


We implemented the proposed performance evaluation in C++ with the
Boost Graph Library (BGL) as a high-quality and high-performance
multithreaded program under the Debian GNU/Linux operating system.

\subsection{First scenario}


In the first scenario, for the offered load $l$, we evaluate the
performance of a PON population characterized by probability $r$ that
an ONU is capable of the IC.  When a specific PON is generated, a
given ONU becomes capable of the IC with probability $r$, and
incapable otherwise.


There are 31 populations, each with a different value of $r$ = \{0,
0.001, 0.002, \ldots, 0.009, 0.01, 0.02, \ldots, 0.09, 0.1, 0.2,
\ldots, 1.0\}.  We evaluate the performance of a population for 11
values of $l$ = \{1, 1.1, 1.2, \ldots, 2\}, totaling 341 values of the
population performance.  We evaluate the performance for $1 \le l \le
2$, because the sharing helps for overloaded networks.


The location of the active RNs is given upfront as shown in
Fig.~\ref{f:scenario}.  At the first stage we install a passive RN,
which is typical for the incumbent PONs.  The high \mbox{1:32}
splitting ratio, which is also typical, and possibly long feeder and
distribution fibers may require an active RN, which we install at the
second stage.  There are on average $gs = 9.6$ active RNs required in
a PON, which is about 10\% of all RNs.  \emph{Today active RNs are
already used to extend the PON reach.}  The last-mile fibers are
typically short, and so passive RNs at the third stage suffice.

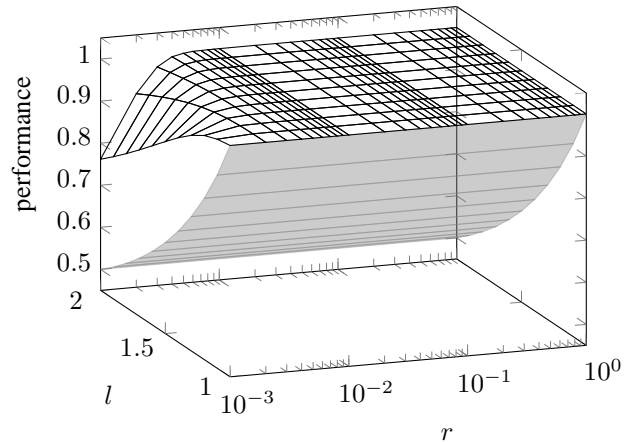
\begin{figure}
  \centering
  \begin{tikzpicture}
  \begin{semilogxaxis}
    [width = 0.9\columnwidth, height = 6.5 cm,
      view/h = -20, view/v = 20,
      xlabel = $r$, ylabel = $l$, zlabel = performance,
      x tick label style = {xshift = 2 pt, yshift = -1 pt},
      y tick label style = {xshift = -3 pt, yshift = 3 pt},
      ztick = {0.5, 0.6, 0.7, 0.8, 0.9, 1}]

    \addplot3 [surf, faceted color = gray, fill = gray, opacity = 0.40] coordinates {
      (0.001, 1, 1)
      (1, 1, 1)

      (0.001, 1.1, 1 / 1.1)
      (1, 1.1, 1 / 1.1)

      (0.001, 1.2, 1 / 1.2)
      (1, 1.2, 1 / 1.2)

      (0.001, 1.3, 1 / 1.3)
      (1, 1.3, 1 / 1.3)

      (0.001, 1.4, 1 / 1.4)
      (1, 1.4, 1 / 1.4)

      (0.001, 1.5, 1 / 1.5)
      (1, 1.5, 1 / 1.5)

      (0.001, 1.6, 1 / 1.6)
      (1, 1.6, 1 / 1.6)

      (0.001, 1.7, 1 / 1.7)
      (1, 1.7, 1 / 1.7)

      (0.001, 1.8, 1 / 1.8)
      (1, 1.8, 1 / 1.8)

      (0.001, 1.9, 1 / 1.9)
      (1, 1.9, 1 / 1.9)

      (0.001, 2, 1 / 2)
      (1, 2, 1 / 2)};

    \addplot3 [mesh, draw = black] table {1st.txt};

  \end{semilogxaxis}
\end{tikzpicture}
  \caption{Performance in the \nth{1} evaluation scenario.}
  \label{f:1st}
\end{figure}

Fig.~\ref{f:1st} shows as the white mesh surface the population
performance with the proposed sharing as a function of $r$ and $l$.
There are 341 data points obtained from 102300 PON performance values
(11 values of offered load $\times$ 31 populations $\times$ 300 PONs
in a population sample).  For comparison, we also plot as the gray
surface the PON performance without the proposed sharing as a function
of $l$ only, since $r$ is irrelevant.

The results show the \emph{proposed sharing improves the PON
performance twice} for the values of $r$ as small as $3 \times
10^{-3}$, which for a PON with $N \approx 3187$ ONUs translates to
about ten IC-ONUs.  For $r = 10^{-3}$, which translates to about only
three IC-ONUs, the PON performance improves from 50\% to 100\%
depending on the offered load.  The PON performance without sharing
caves in as the offered load increases.


For $r = 0$, the sharing cannot take place as there are no IC-ONUs,
and so the white and gray surfaces should meet, but it cannot be
plotted in the logarithmic scale in Fig.~\ref{f:1st}.

\subsection{Second scenario}


In the second scenario, for the offered load $l = 2$, we evaluate the
performance of a PON population described by the probability $r$ that
an ONU is capable of the IC, and by the probability $q$ that an RN is
active.  When a specific PON is generated, a given RN becomes active
with probability $q$, and passive otherwise.


For a given population, the location of active RNs is not fixed, but
is random instead.  For $q = 0$, there are no active RNs, and so there
should be no performance improvement, as the active RNs are
indispensable for the proposed sharing.  For $q = 1$, all RNs are
active.


We have 341 populations, since there are 31 values of $r$ = \{0,
0.001, 0.002, \ldots, 0.009, 0.01, 0.02, \ldots, 0.09, 0.1, 0.2,
\ldots, 1.0\}, and 11 values of $q$ = \{0, 0.1, \ldots, 1\}.  We are
interested in how the proposed sharing improves performance, when a
PON is offered a load twice its capacity, and therefore for each
population we evaluate the performance only for a single value of the
offered load $l = 2$.

\begin{figure}
  \centering
  \begin{tikzpicture}
  \begin{semilogxaxis}
    [width = 0.9\columnwidth, height = 6.5 cm,
    view/h = -20, view/v = 20,
    xlabel = $r$, ylabel = $q$, zlabel = performance,
    x tick label style = {xshift = 2 pt, yshift = -1 pt},
    y tick label style = {xshift = -3 pt, yshift = 3 pt},
    ztick = {0.5, 0.6, 0.7, 0.8, 0.9, 1}]

    \addplot3 [surf, faceted color = gray, fill = gray, opacity =
    0.40] coordinates
    {
    (0.001, 0, 0.5) (0.001, 1, 0.5)

    (1, 0, 0.5) (1, 1, 0.5)
    };

    \addplot3 [mesh, draw = black] table {2nd.txt};

  \end{semilogxaxis}
\end{tikzpicture}
  \caption{Performance in the \nth{2} evaluation scenario.}
  \label{f:2nd}
\end{figure}
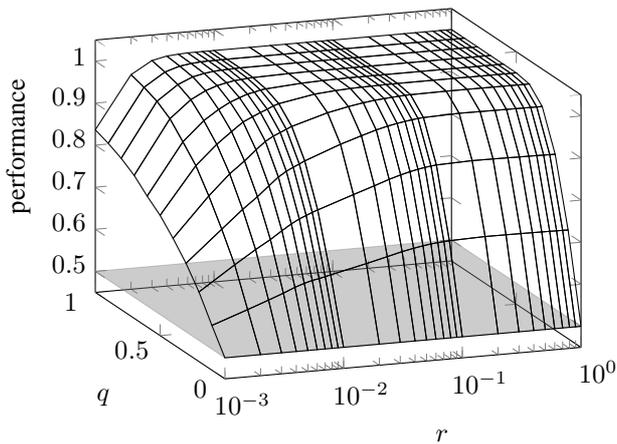


Fig.~\ref{f:2nd}, shows as the white mesh surface the population
performance with the proposed sharing as a function of $r$ and $q$.
There are 341 data points obtained from 102300 PON performance values
(1 value of offered load $\times$ 341 populations $\times$ 300 PONs in
a population sample).  For comparison, we also plot as the gray
surface the PON performance without the proposed sharing.


The results show that the proposed sharing improves the PON
performance, but not as impressively as in the \nth{1} scenario.  For
instance, in the \nth{1} scenario, about 10\% strategically-placed
active RNs at the second stage and about 1\textperthousand{} IC-ONUs
suffice to improve the PON performance about 50\%, while in this
scenario 10\% randomly-placed active RNs and also about
1\textperthousand{} IC-ONUs improve the PON performance about only
10\%.


The performance improvement in this scenario is worse than in the
\nth{1} scenario, because the active RNs which end up at the third
stage are not as useful as they would be, had they been placed at the
second stage as in the \nth{1} scenario, where they would be able to
service more ONUs. 


\section{Conclusion}
\label{conclusion}


We evaluated the performance of the interoperator fixed-mobile network
sharing in the context of the passive optical networks.  The results
suggest that the performance can be improved twofold by providing
interoperator communication to as little as 1\textperthousand{} of all
optical network units.


In the optimistic case, no new hardware is required.  If active remote
nodes are already installed, their software upgrade could be
sufficient to implement the proposed sharing.  Interoperator
communication, in turn, could be software-defined and implemented
wirelessly between base stations.  In the pessimistic case, the
proposed sharing would need the installation of a few
strategically-placed active remote nodes, and providing the
interoperator communication with fiber to a few optical network units.
The proposed sharing is amenable to the pay-as-you-grow deployment,
where the active nodes and intercommunicating optical network units
are deployed in stages when and where needed.


Future work could concentrate on
\begin{inparaenum}
\item extending the dynamic bandwidth allocation protocol to allow for
  the sharing,
\item studying the sharing rules, or
\item generalizing the sharing to any number of operators.
\end{inparaenum} 

\section*{Acknowledgments}

This work was supported by the postdoctoral fellowship number
DEC-2013/08/S/ST7/00576 from the Polish National Science Centre.  The
numerical results were obtained using PL-Grid, the Polish
supercomputing infrastructure.

\bibliographystyle{IEEEtran}
\bibliography{all}

\end{document}